\documentclass[
preprint,
showpacs,
nobibnotes,
 amsmath,amssymb,
 aps,
pra,
floatfix,
]{revtex4-1}

\usepackage{latexsym} 
\usepackage{bm}
\usepackage{graphicx}
\usepackage[usenames]{color} 
\usepackage{dcolumn}

\begin{document}


\title{Hydrodynamic Description of Spin-1 Bose-Einstein Condensates}

\author{Emi Yukawa}
\affiliation{%
 National Institute for Informatics, 2-1-2 Hitotsubashi, Chiyoda-ku, Tokyo 101-8430, Japan
}%


\author{Masahito Ueda}
\affiliation{
 Department of Physics, University of Tokyo, 7-3-1 Hongo, Bunkyo-ku, Tokyo 113-0033, Japan
}%


\date{\today}

\begin{abstract}
We establish a complete set of hydrodynamic equations for a spin-1 Bose-Einstein condensate (BEC), 
which are equivalent to the multi-component Gross-Pitaevskii equations and 
expressed in terms of only observable physical quantities: the spin density and the nematic (or quadrupolar) 
tensor in addition to the density and the mass current that appear in the hydrodynamic description of a scalar BEC. 
The obtained hydrodynamic equations involve a generalized Mermin-Ho relation that is valid regardless of the 
spatiotemporal dependence of the spin polarization. 
Low-lying collective modes for phonons and magnons are reproduced by linearizing the hydrodynamic equations. 
We also apply the single-mode approximation to the hydrodynamic equations and find a complete set of analytic solutions. 

\end{abstract}

\maketitle


\section{\label{sec:1}Introduction}
The standard mean-field description of scalar Bose-Einstein condensates (BECs) is given by the Gross-Pitaevskii (GP) 
equation~\cite{Gross,Pitaevskii} which has proved to be quite instrumental in describing various properties of the system 
such as collective modes~\cite{BaymPethick,Edwards,Cornell,Ketterle} and dynamical instabilities~\cite{Tsubota,Madison1,Madison2,Mottonen,Shin}. 
The hydrodynamic theory provides an equivalent yet intuitively appealing description of the system, since the equations of motion 
are expressed in terms of physical quantities such as the density of particles and the mass current~\cite{Stringari,Dalfovo}. 
The mean-field theory of spinor BECs has been developed based on multi-component GP equations by Ho~\cite{Ho} and 
Ohmi and Machida~\cite{Ohmi}. 
It is natural to ask what are the corresponding hydrodynamic equations and what physical quantities other than the density of 
particles and the mass current are needed to make a complete mean-field description of the spinor BECs. 
Spin domain formation and texture in a quenched spin-1 BEC were observed~\cite{Sadler,Vengalattore} by in-situ and high-resolution 
imaging technique for magnetization profiles for spinor BECs~\cite{Mueller,Higbie}, 
which prompted the hydrodynamic description for ferromagnetic spinor BECs~\cite{Lamacraft1,Kudo1,Kudo2}. 
However, the Berkeley experiment~\cite{Sadler} shows that the magnetization is not fully polarized over the entire condensate. 
Meanwhile, the Majorana representation has been employed to describe general spin states~\cite{Barnett,Lamacraft2}. 
Barnett \textit{et. al.}~\cite{Barnett} have developed the mean-field hydrodynamic equations that involve the Landau-Lifshitz 
equations for the spin-node vectors and reproduce collective excitations from the viewpoint of the 
point-group symmetry, while Lamacraft~\cite{Lamacraft2} has derived the low-energy Lagrangian and obtained the spin-wave spectra. 
In this paper, we derive the most general mean-field hydrodynamic equations for spin-1 BECs that are equivalent to the multi-component GP 
equations and expressed in terms of observable quantities such as the 
magnetization vector and the nematic (or quadrupolar) tensor. 

We consider a BEC of spin-1 bosons which interact via an $s$-wave contact interaction. 
As shown Sec~\ref{sec:3}, a general mean-field state of a spin-1 BEC is completely characterized by the spin and the nematic (or quadrupolar) tensor in 
addition to the density and mass current. 
The density $\hat{\rho}$, the spin $\hat{\bm{F}}$, and the nematic tensor ${\hat{N}}_{\mu \nu}$ are defined as 
\begin{equation} 
	\hat{\rho} \equiv {\hat{\psi}}_{\mu}^{\dagger} {\hat{\psi}}_{\mu}, \label{eq:1-1p-1-1} 
\end{equation} 
\begin{equation} 
	\hat{\bm{F}} \equiv {\hat{\psi}}_{\mu}^{\dagger} {(\bm{F})}_{\mu \nu} {\hat{\psi}}_{\nu}, \label{eq:1-1p-1-2}
\end{equation} 
\begin{equation} 
	{\hat{N}}_{\mu \nu} \equiv {\hat{\psi}}_{\lambda}^{\dagger} {(N_{\mu \nu})}_{\lambda \eta} {\hat{\psi}}_{\eta}, 
	\label{eq:1-1p-1-3} 
\end{equation} 
where a field operator ${\hat{\psi}}_{\mu}$ annihilates a boson with spin index $\mu$, $\bm{F}$ is the vector of spin-1 matrices, and 
$N_{\mu \nu}$ 
is a rank-2 symmetric tensor defined by 
\begin{equation} 
	N_{\mu \nu} 
	= \frac{1}{2} \left ( {(F_{\mu})}_{\lambda \tau} {(F_{\nu})}_{\tau \eta} + {(F_{\nu})}_{\lambda \tau} {(F_{\mu})}_{\tau \eta} 
	\right ). \label{eq:1-1p-2}
\end{equation} 
Throughout this paper, we employ the Cartesian representation and assume that repeated indices are to be summed over the 
Cartesian coordinates (see Sec.~\ref{sec:2}). 
The nematic tensor, which was originally introduced as an order parameter of liquid crystals, has attracted renewed interest in the 
field of ultracold atomic gases as a probe of the nematic order~\cite{Mueller} and in experiments of spin-nematic 
squeezing in an SU(3) system~\cite{Sau,Chapman}. 

The second-quantized Hamiltonian of the system is given by 
\begin{equation}
	\hat{H} = \hat{T} + \hat{U} + {\hat{H}}_{\mathrm{Z}} + \hat{V}, \label{eq:1-1}
\end{equation} 
where $\hat{T}$, $\hat{U}$, ${\hat{H}}_{\mathrm{Z}}$, and $\hat{V}$ represent the kinetic-energy operator, spin-independent one-body potential, 
Zeeman terms, and two-body interaction, which are given as follows: 
\begin{equation}
	\hat{T} = \int d\bm{r} {\hat{\psi}}_{\mu}^{\dagger} \left ( - \frac{{\hbar}^2}{2M} {\nabla}^2 \right ) 
	{\hat{\psi}}_{\mu}, \label{eq:1-2}
\end{equation} 
\begin{equation}
	\hat{U} = \int d\bm{r} U(\bm{r}) {\hat{\psi}}_{\mu}^{\dagger} {\hat{\psi}}_{\mu}, \label{eq:1-3}
\end{equation} 
\begin{equation}
	{\hat{H}}_{\mathrm{Z}} = \int d\bm{r} {\hat{\psi}}_{\mu}^{\dagger} {\left ( -p \bm{n} \cdot \bm{F} + 
	q {\left ( \bm{n} \cdot \bm{F} \right ) }^2 \right )}_{\mu \nu} {\hat{\psi}}_{\nu}, \label{eq:1-4}
\end{equation} 
\begin{equation}
	\hat{V} = \int d\bm{r} \left ( c_0 :{\hat{\rho}}^2: + c_1 :{\hat{\bm{F}}}^2: \right ). \label{eq:1-5}
\end{equation} 
Here $M$ is the mass of a boson, $U(\bm{r})$ is a spin-independent potential such as an optical confinement trap, 
$p$ and $q$ are the coefficients of the linear and quadratic Zeeman energies, respectively, 
$c_0$ and $c_1$ represent the 
spin-independent and spin-exchange interaction energies, and the normal order of operator $\hat{A}$ is denoted as $:\hat{A}:$. 

The mean-field dynamics of the system described by Hamiltonian~(\ref{eq:1-1}) 
is governed by the time-dependent multi-component GP equations: 
\begin{equation} \begin{split}
	i\hbar \frac{\partial {\psi}_{\mu}}{\partial t} =
	\left [ - \frac{{\hbar}^2}{2M} {\nabla}^2 + U(\bm{r}) \right ] {\psi}_{\mu} 
	&+ \left [  -p {(F_z)}_{\mu \nu} + q {(N_{zz})}_{\mu \nu} \right ] {\psi}_{\nu} \\ 
	&+ \left [ c_0 {\delta}_{\mu \nu} + c_1 f_{\lambda} {(F_{\lambda})}_{\mu \nu} \right ] \rho {\psi}_{\nu}. 
	\label{eq:1-8}
\end{split} \end{equation} 
where ${\psi}_{\mu}$ represents the order parameter of the condensate, ${\delta}_{\mu \nu}$ is the Kronecker's delta, and $\rho$ 
and $f_{\mu}$ represent the particle-number density and the spin density: 
\begin{equation} 
	\rho \equiv {\psi}_{\mu}^* {\psi}_{\mu}, \label{eq:1-8p-1-1}
\end{equation}
\begin{equation} 
	f_{\mu} \equiv {\zeta}_{\nu}^* {(F_{\mu})}_{\nu \lambda} {\zeta}_{\lambda}, \label{eq:1-8p-1-2}
\end{equation} 
with ${\zeta}_{\mu}$ being the normalized condensate wave function defined by 
\begin{equation}
	{\zeta}_{\mu} \equiv {\psi}_{\mu} /\sqrt{\rho}. \label{eq:1-8p-1-3} 
\end{equation}  
In the following sections, we will derive a set of hydrodynamic equations that are equivalent to Eq. (\ref{eq:1-8}), and demonstrate 
how to reproduce various properties of the spinor BEC from them. 

This paper is organized as follows. Section~\ref{sec:2} derives the hydrodynamic equations and a generalized Mermin-Ho relation of spin-1 
BECs. 
Section~\ref{sec:3} shows the equivalence of the derived hydrodynamic equations to the multi-component GP equations.
Section~\ref{sec:4} and~\ref{sec:5} are devoted to applications of the obtained hydrodynamic equations: low-lying collective modes are reproduced in a 
physically transparent manner in Sec.~\ref{sec:4}, and the single-mode approximation is employed to the hydrodynamic equations 
to obtain a complete set of analytic solutions in Sec.~\ref{sec:5}. 
Finally, Sec.~\ref{sec:6} summarizes the main results of this paper. 

\section{\label{sec:2}Hydrodynamic equations of spin-1 BECs} 
We adopt the Cartesian basis  $\{ \left | \mu \right >\} =: \mathcal{C}$ ($\mu = x,y,z$) to express 
operators in terms of their matrix representations. 
Each element of $\mathcal{C}$ satisfies 
\begin{equation}
	{\hat{F}}_{\mu} \left | \mu \right > = 0. \label{eq:2-1}
\end{equation} 
The spin matrices in the Cartesian representation are given by 
\begin{equation}
	{(F_{\mu})}_{\nu \lambda} = -i {\epsilon}_{\mu \nu \lambda}, \label{eq:2-3}
\end{equation} 
where ${\epsilon}_{\mu \nu \lambda}$ is the completely antisymmetric unit tensor of rank three. 
Then, the matrix elements of the nematic tensor defined in Eq.~(\ref{eq:1-1p-2}) reduces to 
\begin{equation}
	{(N_{\mu \nu})}_{\lambda \eta} = {\delta}_{\mu \nu} {\delta}_{\lambda \eta} 
	- \frac{1}{2} \left ( {\delta}_{\mu \lambda} {\delta}_{\nu \eta} 
	+ {\delta}_{\nu \lambda} {\delta}_{\mu \eta} \right ). \label{eq:2-4}
\end{equation} 

The hydrodynamic equations for spin-1 BECs are written down in terms of density $\rho$, 
mass current $\bm{v}$, spin density $f_{\mu}$, and nematic tensor $n_{\mu \nu}$, where the mass current and the nematic tensor 
are defined as 
\begin{equation}
	\bm{v} \equiv \frac{\hbar}{2Mi} \left [ {\zeta}_{\mu}^* \left ( \nabla {\zeta}_{\mu} \right ) - 
	\left ( \nabla {\zeta}_{\mu}^* \right ) {\zeta}_{\mu} \right ], \label{eq:2-6} 
\end{equation} 
and 
\begin{equation}
	n_{\mu \nu} \equiv {\zeta}_{\lambda}^* {(N_{\mu \nu})}_{\lambda \eta} {\zeta}_{\eta}, 
	\label{eq:2-8}
\end{equation} 
respectively. 
These variables $\rho$, $\bm{v}$, $f_{\mu}$ and $n_{\mu \nu}$ are referred as the hydrodynamic variables in this paper. 
The spin and nematic currents are defined as: 
\begin{equation}
	{\bm{v}}_{\mu} \equiv \frac{\hbar}{2Mi}
	{(F_{\mu})}_{\nu \lambda} \left [ {\zeta}_{\nu}^* \left (\nabla {\zeta}_{\lambda} \right ) 
	- \left (\nabla {\zeta}_{\nu}^* \right ) {\zeta}_{\lambda} \right ], \label{eq:2-11}
\end{equation} 
\begin{equation}
	{\bm{v}}_{\mu \nu} \equiv \frac{\hbar}{2Mi}
	{(N_{\mu \nu})}_{\lambda \eta} \left [ {\zeta}_{\lambda}^* \left (\nabla {\zeta}_{\eta} \right ) 
	- \left (\nabla {\zeta}_{\lambda}^* \right ) {\zeta}_{\eta} \right ]. \label{eq:2-14}
\end{equation} 
We shall express them in terms of the hydrodynamic variables by making use of the following identity: 
\begin{equation} 
	{\zeta}_{\mu}^* {\zeta}_{\nu} = {\delta}_{\mu \nu} - n_{\mu \nu} + \frac{i}{2} {\epsilon}_{\mu \nu \lambda} f_{\lambda}. 
	\label{eq:2-14p-1}  
\end{equation} 

We derive the hydrodynamic equations by substituting the time-dependent multi-component 
GP equation into the time derivatives of the variables $\rho$, $\bm{v}$, $f_{\mu}$ 
and $n_{\mu \nu}$. 
Firstly, the time derivative of the density $\rho$ leads to the mass continuity equation, i.e., 
\begin{equation}
	\frac{\partial \rho}{\partial t} + \nabla \cdot \rho \bm{v} = 0, \label{eq:2-9}
\end{equation} 
which takes the same form as the mass continuity equation for scalar BECs and ferromagnetic BECs. 

Secondly, we obtain the continuity equation for the spin density as follows: 
\begin{equation}
	\frac{\partial \rho f_{\mu}}{\partial t} + \nabla \cdot \rho {\bm{v}}_{\mu} = 
	\frac{1}{\hbar} {\epsilon}_{z \mu \nu} \rho \left ( p f_{\nu} - 2q n_{z\nu} \right ). \label{eq:2-10}
\end{equation} 
The spin current ${\bm{v}}_{\mu}$ can be expressed in terms of the hydrodynamic variables as follows: 
\begin{equation}
	{\bm{v}}_{\mu} = f_{\mu} \bm{v} - \frac{\hbar}{M} {\epsilon}_{\mu \nu \lambda} \left [ \frac{1}{4} f_{\nu} \left ( \nabla f_{\lambda} \right ) 
	+ n_{\nu \eta} \left ( \nabla n_{\lambda \eta} \right ) \right ], 
	\label{eq:2-12}
\end{equation} 
where the first term on the right-hand side arises from the drift of atoms with the spin density $\bm{f}$, the second and the last terms 
describe the spin currents driven by the spatial variations (i.e., textures) of spin and spin nematicity, respectively. 
The last term in Eq.~(\ref{eq:2-12}) vanishes in the case 
of a fully-polarized (or ferromagnetic) BEC. 
In the general spin-1 case, however, spinor properties are described not only by the spin density but also by the nematic 
tensor as detailed in the next section. 
We also note that Eq.~(\ref{eq:2-10}) reduces to what was obtained in Ref.~\cite{Lamacraft1,Kudo1,Kudo2} in the limit of the fully-polarized 
state, i.e., $|\bm{f} (t)|=1$. 

Thirdly, the continuity equation for the nematic tensor, which does not appear in the case of a ferromagnetic BEC, is given by 
\begin{equation} \begin{split}
	\frac{\partial \rho n_{\mu \nu}}{\partial t} + \nabla \cdot \rho {\bm{v}}_{\mu \nu} 
	= \frac{1}{\hbar} \rho \biggl [ {\epsilon}_{z \mu \lambda} \left ( p n_{\nu \lambda} - \frac{q}{2} {\delta}_{z\nu} f_{\lambda} \right ) 
	&+ {\epsilon}_{z \nu \lambda} \left ( p n_{\mu \lambda} - \frac{q}{2} {\delta}_{z\mu} f_{\lambda} \right ) \biggr ] \\
	&+ \frac{c_1}{\hbar} {\rho}^2 \left ( {\epsilon}_{\mu \lambda \eta} f_{\lambda} n_{\nu \eta} + {\epsilon}_{\nu \lambda \eta} 
	f_{\lambda} n_{\mu \eta} \right ), \label{eq:2-13} 
\end{split} \end{equation} 
with the nematic current ${\bm{v}}_{\mu \nu}$ given by 
\begin{equation}
	{\bm{v}}_{\mu \nu} = n_{\mu \nu} \bm{v} - \frac{\hbar}{4M} \left \{ 
	{\epsilon}_{\mu \lambda \eta} \left [ f_{\lambda} \left ( \nabla n_{\nu \eta} \right ) 
	- \left ( \nabla f_{\lambda} \right ) n_{\nu \eta} \right ] 
	+ {\epsilon}_{\nu \lambda \eta} \left [ f_{\lambda} \left ( \nabla n_{\mu \eta} \right ) 
	- \left ( \nabla f_{\lambda} \right ) n_{\mu \eta} \right ] \right \}. \label{eq:2-15} 
\end{equation} 
The last two terms on the right-hand side of Eq.~(\ref{eq:2-13}) act as spin torques on the nematic tensor, for they can be rewritten as 
\begin{equation}
	\frac{c_1}{\hbar} {\rho}^2 \left ( {\left (\bm{f} \times {\bm{n}}_{\nu} \right )}_{\mu} 
	+ {\left (\bm{f} \times {\bm{n}}_{\mu} \right )}_{\nu} \right ), \label{eq:2-15p-1}
\end{equation}
where ${\bm{n}}_{\mu} \equiv {(n_{\mu x} ,n_{\mu y} ,n_{\mu z})}^T$. 
The nematic current given in Eq.~(\ref{eq:2-15}) involves the texture of this torque force in addition to the mass current associated 
with spin nematicity. 
We will discuss the effect of the torque force in Sec.~\ref{sec:5}. 

Finally, the equation of motion for the mass current is given by: 
\begin{equation} 
\begin{split}
	\frac{\partial v_i}{\partial t} + \left ( v_j {\nabla}_j \right ) v_i 
	&- \frac{{\hbar}^2}{2M^2} {\nabla}_i \frac{{\nabla}_j^2 \sqrt{\rho}}{\sqrt{\rho}} \\
	&+ \frac{{\hbar}^2}{4M^2 \rho} {\nabla}_j \rho \biggl \{ \frac{1}{2} \left [ \left ( {\nabla}_i f_{\mu} \right ) 
	\left ( {\nabla}_j f_{\mu} \right ) 
	- f_{\mu} \left ( {\nabla}_i {\nabla}_j f_{\mu} \right ) \right ] \\ &+  \left [ \left ( {\nabla}_i n_{\mu \nu} \right ) 
	\left ( {\nabla}_j n_{\mu \nu} \right ) - n_{\mu \nu} \left ( {\nabla}_i {\nabla}_j n_{\mu \nu} \right ) 
	\right ] \biggr \} \\
	=& - \frac{1}{M} \left [ \left ( {\nabla}_i U(\bm{r}) \right ) + c_0 \left ( {\nabla}_i \rho \right ) 
	+ c_1 f_{\mu} \left ( {\nabla}_i \rho f_{\mu} \right ) \right ]. \label{eq:2-16}
\end{split}
\end{equation} 
This equation may be regarded as the Euler equation for the spin-1 BEC. 
On the left-hand side, the first two terms represent the usual material derivative and the third one is the quantum-pressure term; 
the remaining terms show the contributions from the spin and nematic textures. 
The right-hand side shows the force terms arising from the gradients of the one-body potential, the particle density, and the spin 
density, respectively. 

The vorticity, which is the rotation of the mass current, is given by
\begin{equation}
	\nabla \times \bm{v} = \frac{1}{2} f_{\mu} \left ( \nabla \times {\bm{v}}_{\mu} \right ) 
	+ n_{\mu \nu} \left ( \nabla \times {\bm{v}}_{\mu \nu} \right ). \label{eq:2-17}
\end{equation} 
Equation~(\ref{eq:2-17}) may be regarded as a generalized Mermin-Ho relation which is valid for an arbitrary state. 
This result shows that both spin and nematic currents contribute to the vorticity through their rotations. 
In the fully-polarized limit, Eq.~(\ref{eq:2-17}) reduces to: 
\begin{equation} 
	f_{\mu} \left ( \nabla \times {\bm{v}}_{\mu} \right ) = 0, \label{eq:2-17p-1}
\end{equation} 
Equation (\ref{eq:2-17p-1}) is another expression of the Mermin-Ho relation~\cite{MerminHo}: 
\begin{equation} 
	\nabla \times \bm{v} - \frac{\hbar}{2M} {\epsilon}_{\mu \nu \lambda} f_{\mu} 
	\left ( \nabla f_{\nu} \times \nabla f_{\lambda} \right ) = 0, 
	\label{eq:2-17p-2}
\end{equation} 
which can be confirmed by substituting Eq.~(\ref{eq:2-12}) and $n_{\mu \nu} = ({\delta}_{\mu \nu} + f_{\mu} f_{\nu})/2$, which holds for 
fully-polarized BECs, into Eq.~(\ref{eq:2-17p-1}). 

For use in Sec.~\ref{sec:5}, we derive the expression of rewrite the energy functional in terms of hydrodynamic variables. 
We start with the mean-field Hamiltonian corresponding to Eq.~(\ref{eq:1-1}), and make the replacement Eq.~(\ref{eq:1-8p-1-3}). 
Using Eq.~(\ref{eq:2-14p-1}), we obtain 
\begin{equation} 
\begin{split}
	E = \int d\bm{r} \biggl \{ \frac{1}{2} M \rho {\bm{v}}^2 + \frac{1}{2M} \left [ {\left ( {\nabla}_i \sqrt{\rho} \right )}^2 
	+ \frac{\rho}{2} {\left ( {\nabla}_i n_{\mu \nu} \right )}^2 + \frac{\rho}{4} {\left ( {\nabla}_i f_{\mu} \right )}^2 \right ]& \\ 
	- p \rho f_z + q \rho n_{zz} + \frac{c_0}{2} {\rho}^2 + \frac{c_1}{2} {\rho}^2 f_{\mu}^2 \biggr \} &.  \label{eq:2-18}
\end{split} 
\end{equation} 
Here, the terms in Eq.~(\ref{eq:2-18}) can also be derived from the symmetry argument 
similar to the Ginzburg-Landau free energy as follows. 
The system is rotationally invariant except for the linear and the quadratic Zeeman terms so that 
the energy functional should be written as 
\begin{equation} 
	E = \int  d\bm{x} {\left [ \frac{M}{2} \rho {\bm{v}}^2 + {\chi}_{\rho} {\left ( {\nabla}_i \rho \right )}^2 
	+ {\chi}_f {\left ( {\nabla}_i f_{\mu} \right )}^2 + {\chi}_n {\left ( {\nabla}_i n_{\mu \nu} \right )}^2 
	+ {\chi}_{\mathrm{int}} + {\chi}_{\mathrm{int}}^{\prime} {\left ( f_{\mu} \right )}^2 \right ]}, \label{eq:2-18p-1} 
\end{equation} 
where ${\chi}_{\rho}$, ${\chi}_f$, ${\chi}_n$, $ {\chi}_{\mathrm{int}}$, and $ {\chi}_{\mathrm{int}}^{\prime}$ 
are functions of $\rho$. 
Here we assume that there are no terms higher than the second order in hydrodynamic variables and a term of 
the second order in the nematic tensor. 
Since 
\begin{equation} 
	n_{\mu \nu}^2 = - \frac{1}{2} f_{\mu}^2 + 2, \label{eq:2-18p-2}
\end{equation}
a term proportional to $n_{\mu \nu}^2$ is absorbed by ${\chi}_{\mathrm{int}}$ 
and ${\chi}_{\mathrm{int}}^{\prime}$. 
Thus, with proper identification of the coefficients, Eq.~(\ref{eq:2-18p-1}) reduces to Eq.~(\ref{eq:2-18}). 
Since only the invariant properties in spin space are used in deriving Eq.~(\ref{eq:2-18p-1}), it might be used for the 
finite-temperature theory of a spinor BEC. 

\section{\label{sec:3}Completeness of the hydrodynamic equations} 
In the case of scalar BECs, the hydrodynamic equations are written in terms of particle-number density $\rho$ 
and mass current $\bm{v}$, and describe the same dynamics as the GP equation.  
The scalar hydrodynamic equations are therefore complete in this sense. 
It is natural to ask whether the hydrodynamic equations for the spin-1 BECs obtained in the 
preceding section can describe the same complete mean-field dynamics as the multi-component GP equations. 
In this section, we answer this question in the affirmative. 

An arbitrary mean-field state of a spin-1 BEC can be obtained by an Euler rotation of a state which 
is, in general, partially polarized in the $z$ direction. 
Thus, a condensate wave function is determined by six variables, which give the same degrees of freedom as 
the three-component condensate wave function. 
Those new variables involve the particle-number density $\rho$, the phase of the U(1) gauge $\phi$, the Euler angles 
$\alpha$, $\beta$, and $\gamma$ defined in Fig.~\ref{fig:3-1}, 
and the polarization 
parameter $\vartheta$ of a $z$-polarized normalized wave function ${\zeta}_{\nu}^{\parallel z} (\vartheta )$: 
\begin{figure}
\centering
\includegraphics[width=5cm,clip]{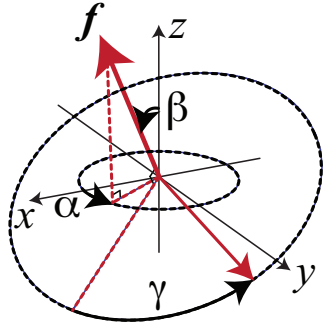}
\caption{Euler angles $\alpha$, $\beta$, and $\gamma$ specifying an arbitrary rotation of the condensate wave function in the 
real space. 
The Euler rotation of a spinor wave function at the coordinate ${\zeta}_{\mu} (\bm{R})$ is described by the matrix representation 
of the rotation operator $\hat{\mathcal{U}}(\alpha (\bm{R}), \beta (\bm{R}), \gamma (\bm{R}))$ given in Eq.~(\ref{eq:3-2}). 
The vector $\bm{f}$ represents the spin vector.} 
\label{fig:3-1}
\end{figure} 
\begin{equation} 
	{\psi}_{\mu} (\alpha ,\beta ,\gamma ; \vartheta )= \sqrt{\rho} {\zeta}_{\mu} (\alpha ,\beta , \gamma ; \vartheta) 
	= \sqrt{\rho} e^{i\phi} {[\mathcal{U} (\alpha ,\beta ,\gamma )]}_{\mu \nu} 
	{\zeta}_{\nu}^{\parallel z} (\vartheta ). \label{eq:3-1} 
\end{equation} 
Here, the matrix ${[\mathcal{U} (\alpha ,\beta ,\gamma )]}_{\mu \nu}$ indicates the Cartesian representation 
of an Euler rotation operator given by 
\begin{equation}
	\hat{\mathcal{U}} (\alpha ,\beta ,\gamma ) = e^{-i{\hat{F}}_z\alpha} e^{-i{\hat{F}}_y\beta} e^{-i{\hat{F}}_z\gamma}, 
	\label{eq:3-2}
\end{equation} 
and ${\zeta}_{\mu}^{\parallel z}$ is written as 
\begin{equation}
	{\zeta}_x^{\parallel z} (\vartheta ) = \cos {\vartheta}, \ {\zeta}_y^{\parallel z} (\vartheta ) = i \sin {\vartheta}, \ 
	{\zeta}_z^{\parallel z} (\vartheta ) = 0, \label{eq:3-3}
\end{equation} 
where the polarization parameter $\vartheta$ parametrizes the three phases of the spin-1 BEC ($n\in \mathbf{Z}$):
\begin{equation}
	\vartheta = \begin{cases} 
	\frac{2n+1}{4} \pi \ \ \ \ \ &(f_{\mu}^2 = F^2 = 1, \text{ fully-polarized or ferromagnetic}); \\
	\frac{n}{2} \pi \ \ \ \ \ &(f_{\mu}^2 = 0, \text{ non-polarized or polar}); \\ 
	\text{otherwise} \ \ \ \ \ &(0<f_{\mu}^2<1, \text{ partially polarized}). \\
	\end{cases} \label{eq:3-4}
\end{equation} 
We shall refer ${[\mathcal{U} (\alpha ,\beta ,\gamma )]}_{\mu \nu} {\zeta}_{\nu}^{\parallel z} (\vartheta )$ 
in Eq.~(\ref{eq:3-1}) as the spinor part of the condensate wave function. 
Then, all the hydrodynamic variables can be rewritten in terms of the six variables of the wave function. 
Here, we examine the spin density $f_{\mu}$, the nematic tensor $n_{\mu \nu}$, and the mass current $\bm{v}$, since 
the density $\rho$ is just the squared magnitude of the wave function. 

\begin{figure}
\centering
\includegraphics[width=4.5cm,clip]{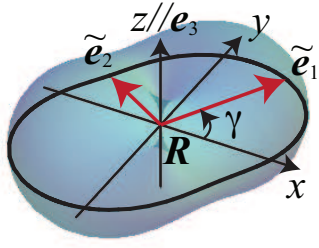}
\caption{Cartesian representation of the condensate wave functions at position $\bm{R}$ plotted in terms of $r(\theta ,\phi )$ given 
in Eq.~(\ref{eq:3-9p-1}) with its two axes ${\tilde{\bm{e}}}_1$ and ${\tilde{\bm{e}}}_2$ indicated by red arrows. 
These two axes are on the plane delimited by closed contour perpendicular to the unit eigenvector of the nematic tensor
${\bm{e}}_3$ at $\bm{R}$. 
For the sake of simplicity, here we set the Euler angles $\alpha =0$ and $\beta = 0$ so that the unit eigenvector of the nematic 
tensor ${\bm{e}}_3$ is parallel to the $z$-axis and $r(\theta ,\phi )= |{\zeta}_{\mu} (0,0, \gamma ;\vartheta ) V_{\mu} (\theta ,\phi )|$. 
The polarization parameter $\vartheta$ is taken to be $\vartheta = 3\pi /16$. 
Note that ${\tilde{\bm{e}}}_1$ and ${\tilde{\bm{e}}}_2$ coincide with the major and minor axes of the condensate wave function, 
respectively.} 
\label{fig:3-2}
\end{figure}

First, we express the spin density $f_{\mu}$ and nematic tensor $n_{\mu \nu}$ in terms of the Euler angles $\alpha$, 
$\beta$, and $\gamma$ and the polarization parameter $\vartheta$. 
The spin density is calculated from Eq.~(\ref{eq:1-8p-1-2}) as 
\begin{equation}
	\bm{f} = \sin {2 \vartheta} {\bm{e}}_f, \label{eq:3-5}
\end{equation} 
where the unit vector ${\bm{e}}_f$ points in the radial direction of the unit vectors of the spherical coordinates, i.e., 
\begin{equation} 
	{\bm{e}}_f = \begin{pmatrix} \cos {\alpha} \sin {\beta} \\ \sin {\alpha} \sin {\beta} \\ \cos {\beta} 
	\end{pmatrix}, \ {\bm{e}}_{\beta} = \begin{pmatrix} \cos {\alpha} \cos {\beta} \\ \sin {\alpha} \cos {\beta} \\
	- \sin {\beta} \end{pmatrix}, 
	{\bm{e}}_{\alpha} = \begin{pmatrix} -\sin {\alpha} \\ \cos {\alpha} \\ 0 \end{pmatrix}. \label{eq:3-6} 
\end{equation} 
In terms of these basis vectors, the nematic tensor given in Eq.~(\ref{eq:2-8}) is diagonalized as: 
\begin{equation}
	n_{\mu \nu} = {\lambda}_r e_{r\mu} e_{r\nu} \ (r \text{ is summed over } r=1,2,3), \label{eq:3-7}
\end{equation} 
with the eigenvalues and eigenvectors given by 
\begin{equation}
	{\lambda}_1 = \frac{1}{2} \left ( 1- \cos {2\vartheta} \right ), \ 
	{\lambda}_2 = \frac{1}{2} \left ( 1+ \cos {2\vartheta} \right ), \ 
	{\lambda}_3 = 1, \label{eq:3-8}
\end{equation} 
and 
\begin{equation} 
	{\bm{e}}_1 = \cos {\gamma} {\bm{e}}_{\beta} + \sin {\gamma} {\bm{e}}_{\alpha}, \ 
	{\bm{e}}_2 = -\sin {\gamma} {\bm{e}}_{\beta} + \cos {\gamma} {\bm{e}}_{\alpha}, \ 
	{\bm{e}}_3 = {\bm{e}}_f. \label{eq:3-9} 
\end{equation} 
Here, the eigenvector ${\bm{e}}_3$ is parallel to the spin vector whose magnitude can be expressed 
in terms of the eigenvalues of the nematic tensor as $2\sqrt{{\lambda}_1 {\lambda}_2}$. 
On the other hand, the remaining eigenvectors ${\bm{e}}_1$ and ${\bm{e}}_2$ are parallel to two of the three principal axes 
of the Cartesian representation of the probability amplitude (see Fig.~\ref{fig:3-2}) given by 
\begin{equation} 
	r (\theta ,\phi ) \equiv \left | {\zeta}_{\mu} (\alpha ,\beta ,\gamma ;\vartheta ) V_{\mu} (\theta ,\phi ) \right |, \label{eq:3-9p-1} 
\end{equation} 
where $V_{\mu}$'s are the basis functions of the Cartesian representation which are defined in terms of the spherical harmonic 
functions $Y^m_l(\theta,\phi)$ as 
\begin{equation} 
\begin{split}
	&V_x (\theta ,\phi ) = \frac{1}{\sqrt{2}} \left ( - Y^1_1 (\theta ,\phi ) + Y^{-1}_1 (\theta ,\phi ) \right ), \\ 
	&V_y (\theta ,\phi ) = \frac{1}{\sqrt{2}} \left ( Y^1_1 (\theta ,\phi ) + Y^{-1}_1 (\theta ,\phi ) \right ), \\
	&V_z (\theta ,\phi ) = Y^0_1 (\theta ,\phi ). \label{eq:3-9p-2} 
\end{split}
\end{equation} 
The axes of the spherical coordinate plot of $r(\theta ,\phi )$ can be written in terms of the eigenvectors of the nematic tensor 
${\bm{e}}_1$ and ${\bm{e}}_2$ as 
\begin{equation}
	{\tilde{\bm{e}}}_r \equiv \sqrt{\frac{3}{8\pi} \left ( 1-{\lambda}_r \right )} \ {\bm{e}}_r \ \ (r=1,2), \label{eq:3-9p-3}
\end{equation} 
which are shown in Fig.~\ref{fig:3-2}. 
These relations between the eigenvectors and eigenvalues of the nematic tensor and the probability amplitudes 
are analogous to the classical electric quadrupole moment that is caused by the distribution of the charge. 

The spin density and nematic tensor have nine independent components, despite the fact that the numbers 
of the variables in the spinor part of the wave function are four, i.e., $\alpha$, $\beta$, $\gamma$, and 
$\vartheta$. 
The degrees of freedom of the spin density and nematic tensor, however, are four which is the same as the 
spinor part of the wave function because of the following five constraints: 
\begin{equation}
	n_{\mu \mu} = 2, \label{eq:3-10}
\end{equation}
\begin{equation}
	n_{\mu \nu} f_{\nu} = f_{\mu}, \label{eq:3-11} 
\end{equation} 
\begin{equation} 
	\mathrm{det} \ n_{\mu \nu} = \frac{1}{4} f_{\mu}^2. \label{eq:3-12}
\end{equation} 

The mass current $\bm{v}$ defined in Eq.~(\ref{eq:2-6}) can be expressed as 
\begin{equation} 
	\bm{v} = \frac{\hbar}{M} \left [ \left ( \nabla \phi \right ) - \left ( \nabla \alpha \right ) \sin {2\vartheta} 
	\cos {\beta} - \left ( \nabla \gamma \right ) \sin {2\vartheta} \right ], \label{eq:3-13} 
\end{equation} 
which can be rewritten in terms of the phase of U(1) gauge $\phi$, the spin vector $f_{\mu}$, and the eigenvectors of the nematic 
tensor $e_{r\mu}$ ($r=1,2,3$) as 
\begin{equation}
	\bm{v} = \nabla \phi - \frac{\hbar}{2M} {\epsilon}_{\mu \nu \lambda} \left [ f_{\mu} e_{r\nu} 
	\left ( \nabla e_{r\lambda} \right ) \right ]. \label{eq:3-14} 
\end{equation} 
Thus, we can determine $\phi$ up to a constant factor from $\bm{v}$ and vice versa, if we know $f_{\mu}$ and 
$n_{\mu \nu}$. 
Here, we note that the rotation of Eq.~(\ref{eq:3-13}) gives another expression of the generalized Mermin-Ho relation. 
Other expressions related to the generalized Mermin-Ho relation have also been discussed in Refs.~\cite{Bouchiat,Hannay,Barnett}. 

We have thus shown that the hydrodynamic variables have the same degrees of freedom as the multi-component condensed wave function for 
a spin-1 BEC, and that a set of the hydrodynamic equations in Eqs.~(\ref{eq:2-9}), (\ref{eq:2-10}), (\ref{eq:2-13}), (\ref{eq:2-16}), and (\ref{eq:2-17}) are 
equivalent to the multi-component GP equations. 

\section{\label{sec:4}Collective modes} 
In this section, low-lying collective modes are derived from the hydrodynamic equations in the absence of an external magnetic field 
in a manner similar to those for a scalar BEC~\cite{Stringari,Dalfovo}. 
We show that the collective modes obtained from the multi-component GP equations with the Bogoliubov approximation are 
fully reproduced in a physically transparent manner. 
\subsection{\label{sec:4-1}Linearization} 
The hydrodynamic equations are linearized as follows. 
The particle-number density $\rho$, the spin density $f_{\mu}$, and the nematic tensor $n_{\mu \nu}$ are decomposed into 
their c-number parts and fluctuations from them as 
\begin{equation} 
	\rho (t,\bm{r}) = \ \bar{\rho} + \delta \rho (t,\bm{r}), 
	f_{\mu} (t,\bm{r}) = {\bar{f}}_{\mu} + \delta f_{\mu} (t,\bm{r}), 
	n_{\mu \nu} (t,\bm{r}) = {\bar{n}}_{\mu \nu} + \delta n_{\mu \nu} (t,\bm{r}). \label{eq:4-1} 
\end{equation} 
The mass current $\bm{v}$ is assumed to be the first order in the fluctuations. 
Then, the second derivatives of the fluctuations of $\rho$ and $f_{\mu}$ with respect to time 
are linearized up to the first order in the fluctuations as follows: 
\begin{equation} 
	\frac{{\partial}^2 \delta \rho}{\partial t^2} = - \frac{{\hbar}^2}{4M^2} \left ( 
	{\nabla}^2 {\nabla}^2 \delta \rho \right ) 
	+ \frac{c_0 + c_1 {\bar{f}}_{\mu}^2}{M} \bar{\rho} \left ( {\nabla}^2 \delta \rho \right ) 
	+ \frac{c_1}{M} {\bar{\rho}}^2 {\bar{f}}_{\mu} \left ( {\nabla}^2 \delta f_{\mu} \right ), 
	\label{eq:4-2} 
\end{equation} 
and
\begin{equation} \begin{split}
	\frac{{\partial}^2 \delta f_{\mu}}{\partial t^2} = - \frac{{\hbar}^2}{4M^2} \left ( {\nabla}^2 {\nabla}^2 
	\delta f_{\mu} \right ) &+ \frac{c_1}{M} {\bar{f}}_{\mu} \left ( 1- {\bar{f}}_{\nu}^2 \right ) \left ( 
	{\nabla}^2 \delta \rho \right ) \\ &+ \frac{c_1}{M} \bar{\rho} \left ( 2 {\bar{n}}_{\mu \nu} - {\delta}_{\mu \nu} 
	- {\bar{f}}_{\mu} {\bar{f}}_{\nu} \right ) \left ( {\nabla}^2 \delta f_{\nu} \right ), \label{eq:4-3}
\end{split} \end{equation} 
where we use the identity 
\begin{equation}
	N_{\mu \lambda} N_{\nu \lambda} = N_{\mu \nu} - \frac{1}{4} 
	\left ( {\delta}_{\mu \nu} f_{\lambda}^2 - f_{\mu} f_{\nu} \right ), \label{eq:4-4}
\end{equation} 
and the constraints in Eqs.~(\ref{eq:3-10})-(\ref{eq:3-12}). 
The linearized equation for the nematic tensor $n_{\mu \nu}$ will be discussed in the next two subsections. 

Collective modes can be decomposed into the density mode and magnon modes, since in the 
linearized modes the first-order terms in fluctuations are taken into account. 
The magnon modes involve two types of modes, i.e., the deformation mode and rotational mode 
with respect to the spin density and the nematic tensor. 
The magnitude of the spin vector or the shape of the nematic tensor varies in the deformation mode, 
while the orientation of the spin vector and the unit eigenvectors of the nematic tensor rotate in the 
rotational mode. 
In the magnon-deformation mode, the polarization $\vartheta$ changes to $\vartheta + \delta \vartheta$, which implies that the 
fluctuations $\delta f_{\mu}$ and $\delta {\lambda}_r$ ($r=1,2$) are given from Eqs.~(\ref{eq:3-5}) and (\ref{eq:3-8}) as: 
\begin{equation}  
	\delta f_{\mu} = 2 \delta \vartheta \cos {2\vartheta}, \label{eq:4-5}
\end{equation} 
\begin{equation} 
	\delta {\lambda}_{1,2} = \pm \delta \vartheta \sin {2\vartheta}. \label{eq:4-6}
\end{equation} 
Those relations will be used in the following subsections. 

\subsection{\label{sec:4-2}Ferromagnetic phase} 
The spin density accompanied by the nematic tensor rotates in the rotational mode that is excited from the ferromagnetic state. 
The spin density and nematic tensor satisfy the following conditions in this case: 
\begin{equation}
	f_{\mu}^2 = 1, {\bar{f}}_{\mu} \delta f_{\mu} = 0, \label{eq:4-1-1}
\end{equation}
\begin{equation}
	n_{\mu \nu} = \frac{1}{2} \left ( {\delta}_{\mu \nu} + f_{\mu} f_{\nu} \right ). \label{eq:4-1-2} 
\end{equation} 
Therefore, the linearized equation of motion for the fluctuation in the spin density is given by 
\begin{equation} 
	\frac{{\partial}^2 \delta f_{\mu} }{\partial t^2} = - \frac{{\hbar}^2}{4M^2} \left ( {\nabla}^2 {\nabla}^2 \delta f_{\mu} \right ). 
	\label{eq:4-1-3}
\end{equation} 
Thus, the dispersion relation of the rotational mode is given by that of a free particle with mass $M$: 
\begin{equation}
	\hbar \omega = \pm {\varepsilon}_{\bm{k}}, \label{eq:4-1-4}
\end{equation} 
where ${\varepsilon}_{\bm{k}}$ denotes the energy of a free particle with mass $M$, i.e., 
${\varepsilon}_{\bm{k}} = {\hbar}^2 k^2 /2M$. 

The deformation mode in the ferromagnetic phase is the mode of the nematic tensor 
(or the spherical plot of the condensate wave function shown in Fig.~\ref{fig:3-2}), 
since the fluctuation of the spin density is second order in the fluctuation of the nematic tensor 
according to Eqs.~(\ref{eq:3-4}), (\ref{eq:3-5}) and (\ref{eq:3-6}). 
The equation of motion for the nematic tensor is linearized as 
\begin{equation} 
	\frac{{\partial}^2 \delta n_{\mu \nu} }{\partial t^2} = - \frac{{\hbar}^2}{4M^2} \left ( {\nabla}^2 
	{\nabla}^2 \delta n_{\mu \nu} \right ) - \frac{2c_1}{M} \bar{\rho} \left ( {\nabla}^2 \delta n_{\mu \nu} \right ) 
	- { \left ( \frac{2c_1}{\hbar} \right )}^2 {\bar{\rho}}^2 \delta n_{\mu \nu}, \label{eq:4-1-7} 
\end{equation} 
which gives the dispersion relation of the deformation mode as 
\begin{equation}
	\hbar \omega = \pm \left | {\varepsilon}_{\bm{k}} - 2 c_1 \bar{\rho} \right |, \label{eq:4-1-8}
\end{equation} 
indicating that a finite gap $2c_1 \bar{\rho}$ arises. 
This mode is interpreted as the mode of $\delta F^2_-$~\cite{Ho}, which is the longitudinal 
spin-wave mode~\cite{Ohmi}, and describes the excitation from $m=1$ to $m=-1$~\cite{Murata}. 

The density mode is obtained from the linearized equation for the 
density fluctuation as 
\begin{equation}
	\frac{{\partial}^2 \delta \rho}{\partial t^2} = - \frac{{\hbar}^2}{4M^2} \left ( {\nabla}^2 {\nabla}^2 
	\delta \rho \right ) - \frac{c_0+c_1}{M} \bar{\rho} \left ( {\nabla}^2 \delta \rho \right ). 
	\label{eq:4-1-9}
\end{equation} 
Hence, the dispersion relation of the density mode is given by 
\begin{equation}
	\hbar \omega = \pm \sqrt{{\varepsilon}_{\bm{k}} \left [ {\varepsilon}_{\bm{k}} 
	+ 2 \left ( c_0 + c_1 \right ) \bar{\rho} \right ]}, \label{eq:4-1-10}
\end{equation} 
which is a Goldstone mode and has the same expression as that for a scalar condensate except for the spin-exchange interaction energy $c_1$. 
This shift is due to the spin-gauge symmetry of the ferromagnetic phase. 

\subsection{\label{sec:4-3}Polar phase}  
In the polar phase, the conditions for the spin density and nematic tensor obey 
\begin{equation}
	f_{\mu} = 0, \label{eq:4-2-1}
\end{equation}
and 
\begin{equation}
	n_{\mu \lambda} n_{\lambda \eta} = n_{\mu \nu}. \label{eq:4-2-2} 
\end{equation} 
Then, the linearized equation of motion for the fluctuation in the nematic tensor is obtained as: 
\begin{equation}
	\frac{{\partial}^2 \delta n_{\mu \nu}}{\partial t^2} = - \frac{{\hbar}^2}{4M^2} \left ( {\nabla}^2 {\nabla}^2 
	\delta n_{\mu \nu} \right ) + \frac{c_1}{M} \bar{\rho} \left ( {\nabla}^2 \delta n_{\mu \nu} \right ). 
	\label{eq:4-2-3}
\end{equation} 
Hence the dispersion relation is given by 
\begin{equation}
	\hbar \omega = \pm \sqrt{{\varepsilon}_{\bm{k}} \left ( {\varepsilon}_{\bm{k}} + 2c_1 \bar{\rho} \right ) }, 
	\label{eq:4-2-4}
\end{equation} 
which is also a Goldstone mode. 

The deformation mode can be understood as the vibrational mode of the spin density. 
In the polar phase, the spin density and nematic tensor satisfy 
\begin{equation}
	{\bar{f}}_{\mu} = 0, \ \delta f_{\mu} = \pm 2 \delta \vartheta e_{f\mu}, \label{eq:4-2-5}
\end{equation} 
and 
\begin{equation} 
	{\bar{\lambda}}_{1,2} = 1,0 \text{ or } {\bar{\lambda}}_{2,1} = 0,1, \ 
	\delta {\lambda}_1 = \delta {\lambda}_2 = \mathcal{O} ({\delta \vartheta}^2), \ 
	{\lambda}_3 = 1. \label{eq:4-2-6}
\end{equation} 
Thus, the linearized equation is given by 
\begin{equation}
	\frac{{\partial}^2 \delta f_{\mu}}{\partial t^2} = - \frac{{\hbar}^2}{4M^2} \left ( {\nabla}^2 {\nabla}^2 
	\delta f_{\mu} \right ) + \frac{c_1}{M} \bar{\rho} \left ( {\nabla}^2 \delta f_{\mu} \right ). \label{eq:4-2-7} 
\end{equation} 
Hence, the dispersion relation is obtained as 
\begin{equation}
	\hbar \omega = \pm \sqrt{{\varepsilon}_{\bm{k}} \left ( {\varepsilon}_{\bm{k}} + 2c_1 \bar{\rho} \right ) }, 
	\label{eq:4-2-8}
\end{equation} 
which has the same dispersion relation as that of the rotational mode in the polar state. 
The magnetization oscillates at the frequency given in Eq.~(\ref{eq:4-2-8}), which can be 
understood as a consequence of the spin-exchange interaction.  

The two degenerate modes in Eq.~(\ref{eq:4-2-4}) and Eq.~(\ref{eq:4-2-8}) have conventionally been interpreted as linearly-independent 
magnon modes~\cite{Ho,Ohmi,Murata}. 
Our hydrodynamic description reveals that they actually describe the rotation of the nematic tensor 
and the amplitude oscillation of the spin, respectively. 

Finally, the linearized equation for the density mode is obtained by substituting $\bar{f}_\mu=0$ into Eq.~(\ref{eq:4-2}): 
\begin{equation}
	\frac{{\partial}^2 \delta \rho}{\partial t^2} = - \frac{{\hbar}^2}{4M^2} \left ( {\nabla}^2 {\nabla}^2 \delta 
	\rho \right ) + \frac{c_0}{M} \bar{\rho} \left ( {\nabla}^2 \delta \rho \right ). \label{eq:4-2-10} 
\end{equation} 
Hence, the dispersion relation is given by 
\begin{equation}
	\hbar \omega = \pm \sqrt{{\varepsilon}_{\bm{k}} \left ( {\varepsilon}_{\bm{k}} + 2 c_0 \bar{\rho} \right ) }. 
	\label{eq:4-2-11} 
\end{equation} 
The density mode from the polar phase is also a Goldstone mode and has the same form as that of a scalar BEC, 
because the mass current in the polar state is given by the same expression as that of a scalar BEC, i.e., 
$\bm{v} = (\hbar /M) \nabla \phi$. 

\section{\label{sec:5}Single-mode approximation} 
As we have seen in Sec.~\ref{sec:2}, the dynamics of a spin-1 BEC is described by a set of the hydrodynamic equations 
in Eqs.~(\ref{eq:2-9}), (\ref{eq:2-10}), (\ref{eq:2-13}), and (\ref{eq:2-16}). 
In the sinigle-mode approximation, we assume that the hydrodynamic variables are spatially uniform. 
The hydrodynamic equations can then be analytically solved. 
Firstly, the continuity equations for the mass, spin vector, and nematic tensor given in Eqs.~(\ref{eq:2-9}), (\ref{eq:2-10}), and 
(\ref{eq:2-13}) reduce to the following equations: 
\begin{equation} 
	\rho = \mathrm{const.}, \label{eq:5-2}
\end{equation} 
\begin{equation} 
	\frac{\partial f_{\mu}}{\partial t} = {\epsilon}_{z \mu \nu} c \left ( \tilde{p} f_{\nu} - 2 \tilde{q} n_{z \nu} \right ) \ (\mu = x,y), \label{eq:5-3} 
\end{equation} 
\begin{equation}
	f_z = m = \mathrm{const.}, \label{eq:5-4}
\end{equation} 
\begin{equation} 
\begin{split}
	\frac{\partial n_{\mu \nu}}{\partial t} = c \biggl [ {\epsilon}_{z \mu \lambda} \left ( \tilde{p} n_{\nu \lambda} 
	- \frac{\tilde{q}}{2} {\delta}_{z \nu} f_{\lambda} \right ) 
	&+ {\epsilon}_{z \nu \lambda} \left ( \tilde{p} n_{\mu \lambda} - \frac{\tilde{q}}{2} {\delta}_{z \mu} f_{\lambda} \right ) \\
	&+ \left ( {\epsilon}_{\mu \lambda \eta} f_{\lambda} n_{\nu \eta} 
	+ {\epsilon}_{\nu \lambda \eta} f_{\lambda} n_{\mu \eta} \right ) \biggr ] , \label{eq:5-5} 
\end{split}
\end{equation} 
where $c \equiv c_1 \rho /\hbar$, $\tilde{p} = p/c \hbar$, and $\tilde{q} = q/c \hbar$. 
Secondly, according to Eq.~(\ref{eq:2-18}), the energy functional with the spatially uniform variables is written as 
\begin{equation} 
	E_{\mathrm{SMA}} = V c \rho \left [ \tilde{q} n_{zz} + \frac{1}{2} \left ( f_x^2 + f_y^2 \right ) \right ] + \mathrm{const.}, \label{eq:5-6}
\end{equation} 
where $V$ denotes the volume of the system and the term ``$\mathrm{const.}$'' represents the 
time-independent part of the energy functional. 
Thus, we define the quantity $\mathcal{E}$ as 
\begin{equation} 
	\mathcal{E} \equiv \tilde{q} n_{zz} + \frac{1}{2} \left ( f_{\mu}^2 - m^2 \right ), \label{eq:5-7}
\end{equation} 
which remains constant in time. 
Finally, we derive the differential equations for the $zz$-component of the nematic tensor $n_{zz}$ and the Euler angle $\alpha$ 
defined in Fig.~\ref{fig:3-1}. 
The other components of the spin vector and nematic tensor can be calculated according to 
Eqs.~(\ref{eq:5-2})-(\ref{eq:5-5}) from these two variables. 
Here, the linear Zeeman energy $\tilde{p}$ causes only the Larmor precession. 
Thus, we set $\tilde{p} =0$ hereafter, since the system with finite $\tilde{p}$ in the rotating frame of the Larmor precession with the 
frequency ${\omega}_{\bm{B}} = p$ is equivalent to the system with $\tilde{p} =0$. 
Then, the differential equation for  $n_{zz}$ is derived from Eqs.~(\ref{eq:5-2})-(\ref{eq:5-5}) as 
\begin{equation} 
\begin{split}
	{\left ( \frac{\partial n_{zz}}{\partial t} \right )}^2 &= 4c^2 \left \{ - m^2 {\left ( n_{zz} - 1\right )}^2 
	+ 2 \left ( \tilde{q} n_{zz} - \mathcal{E} \right ) \left [ n_{zz}^2 - \left ( \frac{1}{2} \tilde{q} + 1 \right ) n_{zz} 
	+ \frac{1}{2} \mathcal{E} \right ] \right \} \\ 
	&\equiv 4c^2 f (n_{zz}). \label{eq:5-8}
\end{split} 
\end{equation} 
Equation~(\ref{eq:5-8}) is equivalent to the equation for the population of the magnetic quantum nuber $m_z =0$ state derived in 
Refs.~\cite{You1,You2}. 
The equation of motion for $\alpha$ is obtained by calculating the time derivative of $\alpha = {\tan}^{-1} {f_y/f_x}$ as 
\begin{equation} 
	\frac{\partial \alpha}{\partial t} = \frac{2\tilde{q} m \left ( n_{zz} - 1 \right )}{\tilde{q} n_{zz} - \mathcal{E}}. \label{eq:5-9}
\end{equation} 
Substituting Eq.~(\ref{eq:5-8}) into Eq.~(\ref{eq:5-9}), we obtain 
\begin{equation} 
	\frac{\partial \alpha }{\partial n_{zz}} = \frac{m}{c} \left [ \frac{1}{\sqrt{f(n_{zz})}} 
	+ \frac{\mathcal{E}/ \tilde{q} -1}{\left ( n_{zz} - \mathcal{E} / \tilde{q} \right ) \sqrt{f(n_{zz})}} 
	\right ]. \label{eq:5-10}
\end{equation} 
Solving Eq.~(\ref{eq:5-10}), we find that $\alpha$ is written in terms of the linear combination of the first-kind and third-kind 
elliptic functions. 

A complete set of hydrodynamic variables can be analytically derived from Eqs.~(\ref{eq:5-8}), (\ref{eq:5-10}), and an initial 
condition. 
In this paper, we solve these hydrodynamic equations in two special cases for the sake of simplicity. 
One is the case $\tilde{q} =0$, i.e., there is no magnetic field; the other is the case $\tilde{q} \neq 0$ and $m=0$, which describes 
the system with a transverse spin a magnetic field. 
\subsection{\label{sec:5-1}$\tilde{q} = 0$} 
The spin vector $f_{\mu}$ remains constant when $\tilde{q}=0$, and $\mathcal{E}$ in Eq.~(\ref{eq:5-7}) becomes 
\begin{equation}
	\mathcal{E} = \frac{1}{2} \left ( f_{\mu}^2 - m^2 \right ). \label{eq:5-11}
\end{equation} 
The differential equation of $n_{zz}$ in Eq.~(\ref{eq:5-8}) reduces to 
\begin{equation} 
	{\left ( \frac{\partial n_{zz}}{\partial t} \right )}^2 = 4c^2 \left [ - { \left ( n_{zz} - \frac{f_{\mu}^2 + m^2}{2f_{\mu}^2} \right )}^2 
	+ \frac{{\mathcal{E}}^2 \left ( 1-f_{\mu}^2 \right )}{{\left (f_{\mu}^2 \right )}^2} \right ], \label{eq:5-12}
\end{equation} 
and $\alpha = \mathrm{const}.$ 
Equation~(\ref{eq:5-12}) has a unique solution for a given arbitrary initial condition. 
Here, we take the initial condition such that 
\begin{equation} 
	\bm{f} (0) = f_0 \begin{pmatrix} 1 \\ 0 \\ 0 \end{pmatrix}, \ n_{\mu \nu} (0) = \begin{pmatrix} 1 & 0 & 0 \\ 
	0 & \frac{1}{2} \left ( 1 + u_0 \right ) & 0 \\ 0 & 0 & \frac{1}{2} \left ( 1 - u_0 \right ) \end{pmatrix}, \label{eq:5-13} 
\end{equation} 
where the magnitude of the spin vector $f_0 \equiv \sin {2\vartheta (0)}$ and $u_0 \equiv \cos {2\vartheta (0)} = \sqrt{1-f_0^2}$  
satisfies $0 < \vartheta (0) < \pi /4$, and the initial Euler angles $\alpha (0)$, $\beta (0)$, and $\gamma (0)$ 
are set to be $\alpha (0) = 0$, $\beta (0) = \pi /2$, and $\gamma (0) = 0$. 
The solution of the hydrodynamic equations is given by 
\begin{equation} 
	n_{\mu \nu} (t) = \begin{pmatrix} 1 & 0 & 0 \\ 0 & \frac{1}{2} \left ( 1+ u_0 \cos {2\omega t} \right ) & \frac{1}{2} u_0 
	\sin {2\omega t} \\ 
	0 & \frac{1}{2} u_0 \sin {2\omega t} & \frac{1}{2} \left ( 1 - u_0 \cos {2\omega t} \right ), \end{pmatrix}, \label{eq:5-14} 
\end{equation}
with 
\begin{equation} 
	\omega = c |\bm{f}|. \label{eq:5-15}
\end{equation} 
The obtained nematic tensor in Eq.~(\ref{eq:5-14}) is diagonalized as $n_{\mu \nu} = {\lambda}_r e_{\mu r} e_{\nu r}$ (the sum 
is taken over $r=1,2,3$), 
where 
\begin{equation} 
	{\lambda}_1 = \frac{1}{2} \left ( 1+u_0 \right ), \ {\lambda}_2 = \frac{1}{2} \left ( 1-u_0 \right ), \ {\lambda}_3 = 1,  \label{eq:5-16}
\end{equation} 
and 
\begin{equation} 
	{\bm{e}}_1 = \begin{pmatrix} 0 \\ \cos {\omega t} \\ \sin {\omega t} \end{pmatrix}, \ 
	{\bm{e}}_2 = \begin{pmatrix} 0 \\ - \sin {\omega t} \\ \cos {\omega t} \end{pmatrix}, \ 
	{\bm{e}}_3 = \begin{pmatrix} 1 \\ 0 \\ 0 \end{pmatrix}, \label{eq:5-17}
\end{equation} 
which implies that the quadrupole moment rotates about the spin vector at frequency $\omega = c|\bm{f}|$, despite the fact 
that there is no external magnetic field. 
This rotation of the nematic tensor is caused by the source terms on the right-hand side of Eq.~(\ref{eq:5-5}); the spin vector acts 
like a torque force on the nematic tensor because of the spin-exchange interaction. 

This rotational mode of the nematic tensor is not the magnon collective excitation from the partially polarized 
state~\cite{Murata}, but the mode caused by the spin-mixing derived in Ref.~\cite{Pu}.
The magnon collective excitation from the partially polarized state asymptotically approaches the deformation mode excited from 
the ferromagnetic phase derived in Eq.~(\ref{eq:4-1-8}) if we take the limit $\tilde{q} \to 0$ with $c <0$ and 
$| \tilde{q} /\tilde{p}| >0$, while keeping $\tilde{q} /\tilde{p}$ constant. 

Finally, we note that this phenomenon is expected to be observed in the following experimental situation. 
First, a spin-1 BEC is prepared in a partially polarized state, for instance, in a BEC 
that consists of 80\% particles in the $m = 1$ state and 20\% particles in the $m = -1$ state. 
Then, we excite particles in the $m = -1$ ($1$) to the $m = 1$ ($-1$) state by a two-photon Raman process involving absorption of 
${\sigma}_+$ (${\sigma}_-$) light and emission of ${\sigma}_-$ (${\sigma}_+$) light. 
If the frequency difference between two laser beams matches the transition frequency between $m = -1$ and $m = 1$, we can 
expect a resonance phenomenon at $\omega = c|\bm{f}|$. 

\subsection{\label{sec:5-2}$\tilde{q} \neq 0$, $m=0$} 
A physical solution always exists in the case $\tilde{q} \neq 0$ and $m=0$. 
All hydrodynamic variables can be derived from the initial condition and $n_{zz} (t)$, since the angle of the spin vector in the 
$x$-$y$ plane, $\alpha$, which obeys the differential equation given in Eq.~(\ref{eq:5-9}), remains constant. 
Thus, the equation of motion relevant to the dynamics is the differential equation for $n_{zz}$ given in Eq.~(\ref{eq:5-8}), which is simplified as 
\begin{equation}
	{\left ( \frac{\partial n_{zz}}{\partial t} \right )}^2 = 8c^2 \tilde{q} \left ( n_{zz} - \frac{\mathcal{E}}{\tilde{q}} \right ) 
	\left ( n_{zz} - {\lambda}_+ \right ) \left ( n_{zz} - {\lambda}_- \right ), \label{eq:5-18}
\end{equation} 
where two of three zeros, i.e.,  ${\lambda}_{\pm}$, are given by 
\begin{equation}
	{\lambda}_{\pm} \equiv \frac{1}{2} \left [ 1+ \frac{\tilde{q}}{2} \pm \sqrt{{\left ( 1+\frac{\tilde{q}}{2} \right )}^2 
	- 2 \mathcal{E}} \right ]. \label{eq:5-19} 
\end{equation} 
There always exists a physical solution of Eq.~(\ref{eq:5-18}) with respect to an arbitrary initial condition, which can be 
analytically shown in a straightforward manner. 
The $zz$-component of the nematic tensor $n_{zz}$ obtained from Eq.~(\ref{eq:5-18}) is expressed in terms of the first-kind elliptic 
functions with elliptic modulus $k$ determined from $\tilde{q}$ and the initial condition, which implies that the dynamics has the 
period $2T(k)$, where $T(k)$ is the first-kind elliptic integral. 
Unlike the case of $\tilde{q} =0$, the magnitudes of the spin vector $|\bm{f}|$ and the axes of the spherical plot of the condensate 
wave function $|{\tilde{\bm{e}}}_1|$ and $|{\tilde{\bm{e}}}_1|$ defined in Eq.~(\ref{eq:3-9p-3}) (see also Fig.~\ref{fig:3-2}) depend on time. 
The eigenvectors of the nematic tensor ${\bm{e}}_1$ and ${\bm{e}}_2$ rotate due to the torque force caused by the 
spin-exchange interaction as calculated for the case of $\tilde{q} =0$; however, they do not always undergo circular motion, that is 
to say, $\gamma (t)$ does not necessarily range from $-\pi$ to $\pi$, as a consequence of the energy balance between the 
quadratic Zeeman energy $\tilde{q}$ and the spin-exchange energy $c$. 
Equation~(\ref{eq:5-18}) can be solved analytically with the same initial condition as in the case of $\tilde{q}=0$ in Eq.~(\ref{eq:5-13}). 
The solution depends on the condition for $\tilde{q}$ and $u_0$ and the complete solution is listed in Appendix A and some typical 
examples are illustrated in Fig. 3.  

\section{\label{sec:6}Summary} 
In this paper, we have derived a complete set of the hydrodynamic equations for a spin-1 BEC in an arbitrary state, which give a 
description equivalent to the multi-component GP equations. 
The hydrodynamic equations are self-contained within the hydrodynamic variables: the particle-number density, the mass current, 
the spin density, and the nematic tensor, as shown in Sec.~\ref{sec:2}. 
The energy functional in terms of the hydrodynamic variables has also been derived. 
The obtained equations involve the continuity equation for the density, the spin density, and the nematic tensor, the equation 
of motion for the mass current, and a generalized Mermin-Ho relation for arbitrary spin-1 BECs, as discussed 
in Sec.~\ref{sec:3}. 
We have applied our hydrodynamic equations to some specific situations in Secs.~\ref{sec:4} and \ref{sec:5}. 
In Sec.~\ref{sec:4}, we have shown that our hydrodynamic equations reproduce the collective modes obtained by the multi-component GP 
equations~\cite{Ho,Ohmi}.  
In Sec.~\ref{sec:5}, the single-mode approximation have been applied to the continuity equations so that the spin density acts 
on the nematic tensor like a torque as a consequence of the spin-exchange interaction. 
Under a magnetic field, the torque force and the quadratic Zeeman effect balance and the dynamics of BEC can be described 
by the first-kind elliptic functions as derived in Ref.~\cite{You1,You2}. 

\appendix 
\section*{ACKNOWLEDGMENTS} 
We gratefully acknowledge Yusuke Kato and Yuki Kawaguchi for a number of constructive comments and discussions. 
This work was supported by Grants-in-Aid for Scientific Research (Kakenhi Nos. 22340114 and 22103005), a Global 
COE Program ``The Physical Science Frontier'', and the Photon Frontier Netowork Program of MEXT of Japan. 

\section{Solutions of Eq.~(\ref{eq:5-18})} 
\begin{description} 
\item[(i) $\bm{ -1-u_0<\tilde{q}<-2u_0, \ 0<\tilde{q}<1-u_0}$] \mbox{} \\ 
The spin vector is always parallel to the $x$-axis because of the initial condition in Eqs.~(\ref{eq:5-13}) and the constant $\alpha$ 
and written as 
\begin{equation} 
	\bm{f} (t) =  f_0 \ \mathrm{dn} \left ( cf_0t, k \right ) \begin{pmatrix} 1 \\ 0 \\ 0 \end{pmatrix}, \label{eq:5-20}
\end{equation} 
where, $\mathrm{dn} (x,k)$ is defined in terms of the first-kind elliptic function $\mathrm{sn}(x,k)$ as 
\begin{equation} 
	\mathrm{dn} \left (x,k \right ) \equiv \sqrt{1-k^2 {\mathrm{sn}}^2 \left (x,t \right )}, \label{eq:5-21}
\end{equation} 
and $k$ is given by 
\begin{equation} 
	k \equiv \frac{\sqrt{\tilde{q}\left ( 2u_0+\tilde{q}\right )}}{f_0}. \label{eq:5-22}
\end{equation} 
Next, the nematic tensor is written as  
\begin{equation} 
	n_{\mu \nu} (t) = \begin{pmatrix} 1 & 0  & 0 \\ 0 & 1-n_{zz} (t) & n_{yz} (t) \\ 0 & n_{yz} (t) & n_{zz} (t) \end{pmatrix}, \label{eq:5-23}
\end{equation} 
since the nematic tensor, which is a symmetric tensor with the condition in Eq.~(\ref{eq:3-10}), is diagonalized as Eq.~(\ref{eq:3-7}) 
with the eigenvectors in Eqs.~(\ref{eq:3-9}) with ${\bm{e}}_3 = (1,0,0)^T$. 
The components of the nematic tensor are obtained from Eq. (\ref{eq:5-18}) and the $yz$-component in Eq. (\ref{eq:5-5}) as 
\begin{equation} 
	n_{yz} (t) = \frac{1}{2} \left ( 2u_0+\tilde{q} \right ) \mathrm{sn} \left ( cf_0t, k \right ) \mathrm{cn} ( cf_0t, k), \label{eq:5-24}
\end{equation} 
\begin{equation} 
	n_{zz} (t) = \frac{1}{2} \left [ 1-u_0 
	+ \left ( 2u_0 + \tilde{q} \right ) {\mathrm{sn}}^2 \left ( cf_0t, k \right ) \right ]. \label{eq:5-25}
\end{equation} 
Finally, the Euler angle $\gamma (t)$, which represents the rotational angle of two axes of the nematic tensor ${\bm{e}}_1$ and 
${\bm{e}}_2$ about the spin vector, is uniquely determined by requiring that it satisfy 
\begin{equation} 
	\gamma (t) = \frac{1}{2} 
	{\cos}^{-1} {\left [ \frac{u_0 \ {\mathrm{cn}}^2  
	- \left ( u_0 + \tilde{q} \right ) {\mathrm{sn}}^2 }{\sqrt{u_0^2{\mathrm{cn}}^2 
	+ {\left ( u_0 + \tilde{q} \right )}^2{\mathrm{sn}}^2 }} \right ] }, \label{eq:5-26}
\end{equation} 
and 
\begin{equation} 
	\gamma (t) = \frac{1}{2} {\sin}^{-1} {\left [ \frac{\left ( 2u_0 + \tilde{q} \right ) \mathrm{sn}  
	\mathrm{cn} }{\sqrt{u_0^2{\mathrm{cn}}^2 
	+ {\left ( u_0 + \tilde{q} \right )}^2{\mathrm{sn}}^2 }} \right ]}, \label{eq:5-27}
\end{equation} 
where ``$\mathrm{sn}$'' and ``$\mathrm{cn}$'' are the abbreviations of $\mathrm{sn} \left (cf_0t,k \right )$ and $\mathrm{cn} \left (cf_0t,k \right )$, 
respectively. 
Taking the limit $\tilde{q} \to 0^+$, Eqs~(\ref{eq:5-26}) and (\ref{eq:5-27}) reduce to $\gamma (t)$ in Eq.~(\ref{eq:5-15}), that is to 
say,  $\gamma (t) = cf_0t$. 
\item[(ii) $\bm{\tilde{q} = 1-u_0}$] \mbox{} \\ 
The elliptic modulus $k=1$ in the case of $\tilde{q} = 1-u_0$, which implies that the first-elliptic function $\mathrm{sn}$ becomes 
the hyperbolic function $\tanh$. 
The spin vector is given by 
\begin{equation}
	\left |\bm{f} (t) \right | = \frac{f_0}{\cosh {cf_0t}}. \label{eq:5-28}
\end{equation} 
The nematic tensor is written as in Eq.~(\ref{eq:5-23}) with $n_{yz} (t)$ and $n_{zz} (t)$ given by 
\begin{equation} 
	n_{yz} (t) = \frac{1}{2} \left ( 1+ u_0 \right ) \frac{\tanh {cf_0t}}{\cosh {cf_0t}}, \label{eq:5-29}
\end{equation} 
and 
\begin{equation} 
	n_{zz} (t) = \frac{1}{2} \left [ 1-u_0 + \left ( 1+u_0 \right ) {\tanh}^2 cf_0t \right ], \label{eq:5-30}
\end{equation} 
respectively. 
A BEC asymptotically becomes to the polar state in the limit $t \to \infty$: 
\begin{equation} 
\begin{split}
	&\lim_{t\to \infty} \bm{f} (t) = \bm{0}, \\ &\lim_{t\to \infty} n_{\mu \nu} (t) = \begin{pmatrix} 1 & 0 & 0 \\ 0 & 0 & 0 \\ 0 & 0 & 1 
	\end{pmatrix}, \label{eq:5-31} 
\end{split}
\end{equation} 
which implies that the axis of the nematic tensor in the polar state is along the $y$-axis, i.e., the direction perpendicular to the 
magnetic field. 
\item[(iii) $\bm{\tilde{q}>1-u_0, \ \tilde{q} < -1-u_0}$] \mbox{} \\ 
The spin vector is calculated as 
\begin{equation} 
	\bm{f} (t) = f_0 \mathrm{cn} \left (cf_0t/k,k \right ), \label{eq:5-32}
\end{equation} 
where 
\begin{equation}
	k = \frac{f_0}{\sqrt{\tilde{q} \left (2u_0+\tilde{q} \right ) }}. \label{eq:5-33}
\end{equation} 
The nematic tensor has the same form as in Eq.~(\ref{eq:5-23}), and $n_{yz} (t)$ and $n_{zz} (t)$ are written as 
\begin{equation} 
	n_{yz} (t) = \frac{f_0^2}{2\tilde{q}k} \mathrm{sn} (cf_0t/k,k) \mathrm{dn} (cf_0t/k,k), \label{eq:5-34}
\end{equation} 
\begin{equation} 
	n_{zz} (t) = \frac{1}{2} \left ( 1-u_0 + \frac{f_0^2}{\tilde{q}} 
	{\mathrm{sn}}^2 (cf_0t/k,k) \right ). \label{eq:5-35}
\end{equation} 
The angle $\gamma (t)$, which is the rotation angle of the eigenvectors of 
the nematic tensor ${\bm{e}}_1$ and ${\bm{e}}_2$, satisfies 
\begin{equation} 
	\gamma (t) = \frac{1}{2} {\cos}^{-1} \left [ 
	\frac{u_0 {\mathrm{dn}}^2 - \left ( u_0 + \tilde{q} \right ) k^2 {\mathrm{sn}}^2 }{\sqrt{u_0^2 {\mathrm{dn}}^2 
	+{\left ( u_0 + \tilde{q} \right )}^2  k^2 {\mathrm{sn}}^2}}
	\right ],  \label{eq:5-36}
\end{equation} 
and 
\begin{equation} 
	\gamma (t) = \frac{1}{2} {\sin}^{-1} \left [ \frac{f_0^2}{\tilde{q}k} 
	\frac{\mathrm{dn} \ \mathrm{sn} }{\sqrt{u_0^2 {\mathrm{dn}}^2 + {\left ( u_0 + \tilde{q} \right )}^2  k^2 {\mathrm{sn}}^2}}
	\right ],  \label{eq:5-37}
\end{equation} 
where ``$\mathrm{dn}$'' and ``$\mathrm{sn}$'' represent the abbreviations of $\mathrm{dn} (cf_0t/k,k)$ and 
$\mathrm{sn} (cf_0t/k,k)$, respectively. 
\item[(iv) $\bm{-2u_0<\tilde{q}<0}$] \mbox{} \\ 
The spin vector is expressed as 
\begin{equation}
	\bm{f} (t) = f_0 \mathrm{nd} \left (cf_0t/\sqrt{1-k^2}, k \right ) \begin{pmatrix} 1 \\ 0 \\ 0 \end{pmatrix}, \label{eq:5-38}
\end{equation} 
where $\mathrm{nd} (x,k) \equiv 1/\mathrm{dn} (x,k)$ and the elliptic modulus $k$ is given by 
\begin{equation} 
	k = \frac{1}{\sqrt{1 -\frac{f_0^2}{\tilde{q} \left ( 2u_0 + \tilde{q} \right )} }}. \label{eq:5-39}
\end{equation} 
In this case, we have 
\begin{equation} 
	\left | \bm{f} (t) \right | \le 1, \label{eq:5-39p-1}
\end{equation} 
with the equality holding if and only if $\tilde{q} = -u_0$. 
For the sake of simplicity, the arguments $(cf_0t/\sqrt{1-k^2}, k)$ in elliptic functions are not written explicitly in the following 
part of this example. 
The nematic tensor is written as the expression given in Eq.~(\ref{eq:5-23}) with 
\begin{equation} 
	n_{yz} (t) = \frac{f_0^2k^2}{2|\tilde{q}|\sqrt{1-k^2}} \mathrm{cd} \ \mathrm{sd} , \label{eq:5-40}
\end{equation} 
\begin{equation} 
	n_{zz} (t) = \frac{1}{2} \left ( 1-u_0 + \frac{f_0^2k^2}{|\tilde{q}|} {\mathrm{sd}}^2 \right ), \label{eq:5-41}
\end{equation} 
where $\mathrm{cd} \equiv \mathrm{cn} /\mathrm{dn}$ and $\mathrm{sd} \equiv \mathrm{sn} /\mathrm{dn}$. 
The rotation angle $\gamma (t)$ is determined from the following equations: 
\begin{equation} 
	\gamma (t) = \frac{1}{2} {\cos}^{-1} {\left [ \frac{u_0 {\mathrm{cd}}^2 - \left ( u_0 + \tilde{q} \right ) \left ( 1-k^2 \right ) 
	{\mathrm{sd}}^2}{\sqrt{u_0^2 {\mathrm{cd}}^2 
	+ {\left ( u_0 + \tilde{q} \right )}^2 \left ( 1-k^2 \right ) {\mathrm{sd}}^2}} \right ]}, \label{eq:5-42}
\end{equation} 
and 
\begin{equation} 
	\gamma (t) = \frac{1}{2} {\sin}^{-1} \left [  \frac{f_0^2k^2}{|\tilde{q}| \sqrt{1-k^2}} \frac{\mathrm{cd} \ 
	\mathrm{sd}}{\sqrt{u_0^2 {\mathrm{cd}}^2 
	+ {\left ( u_0 + \tilde{q} \right )}^2 \left ( 1-k^2 \right ) {\mathrm{sd}}^2}} \right ], \label{eq:5-43}
\end{equation} 
except for the case of $\tilde{q} = -u_0$, where $\gamma (t)$ is given by the following equations: 
\begin{equation} 
\begin{split}
	& \gamma (t) = \frac{1}{2} {\cos}^{-1} \left [ {\mathrm{cd}} \right ], \\
	& \gamma (t) = \frac{1}{2} {\sin}^{-1}  \left [ {f_0 \mathrm{sd}} \right ]. \label{eq:5-43p-1} 
\end{split}
\end{equation} 
The eigenvectors of the nematic tensor ${\bm{e}}_1$ and ${\bm{e}}_2$ trace closed curves in the case of $-2u_0 < \tilde{q} < -u_0$ 
as shown in (1) of Fig.~\ref{fig:5-1}, 
whereas they rotates about ${\bm{e}}_3$ when $-u_0 \le \tilde{q} < 0$ as an example shown in (3) of Fig.~\ref{fig:5-1}. 
\begin{figure*}
\centering
\includegraphics[width=15cm,clip]{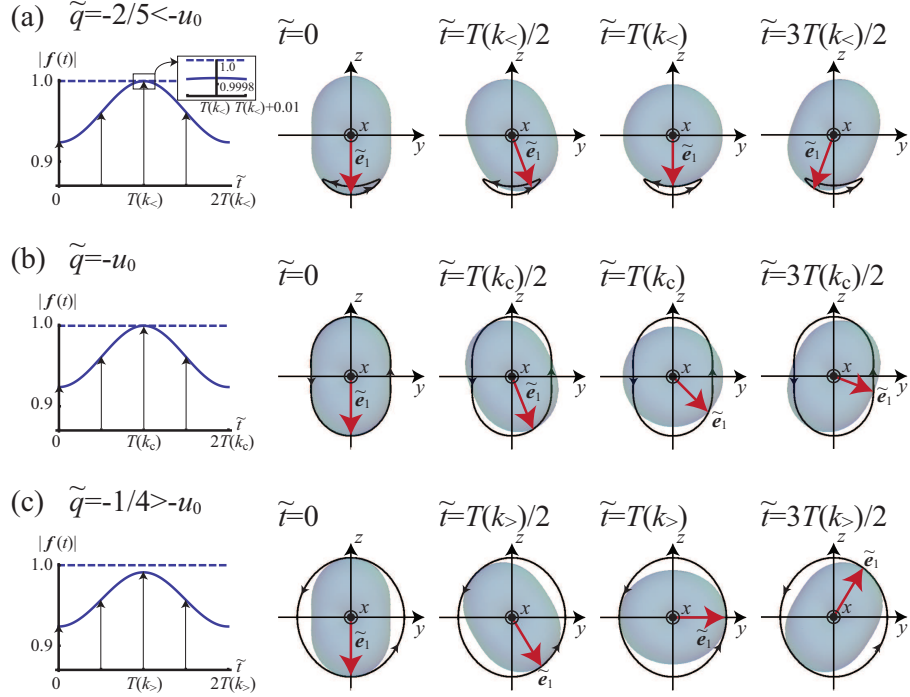}
\caption{Time and $\tilde{q}$ dependences of $|\bm{f} (t)|$ (each leftmost panel) and snapshots of the condensate wave functions 
with their principal axes ${\tilde{\bm{e}}}_1$'s indicated by red arrows for 
the case of (a) $\tilde{q} =-2/5<u_0$, (b) $\tilde{q} = -u_0$, and (c) $\tilde{q} = -1/4 > -u_0$, where $\tilde{q}$'s satisfy 
$-2u_0 < \tilde{q} <0$.  
Here, $\vartheta (0)$ is taken to be $\vartheta (0) = 3\pi /16$ and hence $u_0 = \cos {3\pi /8}$ and $f_0 = \sin {3\pi /8}$. 
The period in each case is given by $2T(k(\tilde{q}))$, where $T(k(\tilde{q}))$ is the first-kind elliptic integral with the elliptic modulus 
$k(\tilde{q})$ given by Eq.~(\ref{eq:5-39}). 
We abbreviate (a) $k_< \equiv k(-2/5)$, (b) $k_c \equiv k(-u_0)$, and (c) $k_> \equiv k(-1/4)$, here. The track of 
${\tilde{\bm{e}}}_1$, which is indicated by solid closed curves with arrows, undergoes a transition at a branch point (b): 
$\tilde{q} = -u_0$. 
The principal axes swing around their initial position for the case of $\tilde{q} < - u_0$ as shown in 
(a), whereas they rotate about the spin vector (or the eigenvector of the nematic tensor ${\bm{e}}_3$) for the case of 
$\tilde{q} > -u_0$ as shown in (c), because for $\tilde{q} = -u_0$, the major and minor axes exchange each other at $t =T(k_c)$, 
where the spin vector is fully 
polarized and the major and minor axes can take any mutually orthogonal directions.}
\label{fig:5-1}
\end{figure*} 
\item[(v) $\bm{\tilde{q}=-2u_0}$] \mbox{} \\ 
This case is trivial, where the spin vector and the nematic tensor remain constant.  
\item[(vi) $\bm{\tilde{q}=-1-u_0}$] \mbox{} \\ 
A BEC asymptotically becomes the polar state in the limit $t \to \infty$. 
The spin vector is obtained as 
\begin{equation} 
	\bm{f} (t) = \frac{f_0}{\cosh {cf_0t}}, \label{eq:5-44}
\end{equation} 
and the nematic tensor, which is expressed as Eq.~(\ref{eq:5-23}), involves the component $n_{yz} (t)$ and $n_{zz} (t)$ given by 
\begin{equation} 
	n_{yz} (t) = - \frac{1}{2} \left ( 1-u_0 \right ) \frac{\tanh {cf_0t}}{\cosh {cf_0t}}, \label{eq:5-45}
\end{equation} 
and 
\begin{equation} 
	n_{zz} (t) = \frac{1-u_0}{2{\cosh}^2 {cf_0t}}, \label{eq:5-46}
\end{equation} 
respectively. 
In the limit $t \to \infty$, the spin vector and nematic tensor grow asymptotically to 
\begin{equation} 
	\lim_{t\to \infty} \bm{f} (t) = \bm{0}, \ \lim_{t\to \infty} n_{\mu \nu} (t) = \begin{pmatrix} 1 & 0 & 0 \\ 0 & 1 & 0 \\ 
	0 & 0 & 0 \end{pmatrix}, 
	\label{eq:5-47}
\end{equation} 
where the axis of the nematic tensor in the polar state is along the $z$-axis, i.e., the direction of the magnetic field. 
\end{description}



\end{document}